\documentclass[aps,preprint,amsmath,amssymb]{revtex4}
\usepackage{graphicx}

\def\hmp{\hat m_{P}}

\def\thl{\theta_{\ell}}

\def\tm{\tilde{m}}

\begin{document}

\title{\Large \bf $\eta^{(\prime)}$ productions in semileptonic B decays}

\author{ \bf \large Chuan-Hung Chen$^{1,2}$\footnote{Email:
physchen@mail.ncku.edu.tw} and Chao-Qiang
Geng$^{3,4}$\footnote{Email: geng@phys.nthu.edu.tw}
 }

\affiliation{ $^{1}$Department of Physics, National Cheng-Kung
University, Tainan 701, Taiwan \\
$^{2}$National Center for Theoretical Sciences, Taiwan\\
$^{3}$Department of Physics, National Tsing-Hua University, Hsinchu
300, Taiwan  \\
$^{4}$Theory Group, TRIUMF, 4004 Wesbrook Mall, Vancouver, B.C V6T 2A3, Canada
 }

\date{\today}

\begin{abstract}
Inspired by the new measurements on $B^{-}\to \eta^{(\prime)}\ell
\bar\nu_{\ell}$ from the BaBar Collaboration, we examine the
constraint on the flavor-singlet mechanism, proposed to understand the
large branching ratios for $B\to \eta^{\prime} K$ decays. Based on
the mechanism, we study the decays of $\bar B_{d,s}\to
\eta^{(\prime)} \ell^{+} \ell^{-}$ and find that they are sensitive
to the flavor-singlet effects. In particular, we show that the decay
branching ratios of $\bar B_{d,s}\to \eta^{\prime} \ell^{+}
\ell^{-}$ can be as large as $O(10^{-8})$ and $O(10^{-6})$,
respectively.

\end{abstract}

\maketitle %

Until now, the unexpected large branching ratios (BRs) for the decays
$B\to \eta^{\prime} K $ are still mysterious phenomena among the
enormous measured exclusive $B$ decays at $B$ factories
\cite{belle_0603,babar_0608}.
One of the most promising mechanisms to understand
the anomaly is to introduce
a flavor-singlet state, produced by the two-gluon emitted from the
light quarks in $\eta^{(\prime)}$ \cite{Kroll,BN_NPB651}.
In this mechanism, the form factors in
the $B\to\eta^{(\prime)}$ transitions receive  leading power
corrections. Consequently,
the authors in Ref.~\cite{KOY} have studied
the implication on the semileptonic decays of $\bar
B\to P\ell \nu_{\ell}$ with $P=\eta^{(\prime)}$ and $\ell=e,\ \mu$.
In particular, they find that the decay BRs of the $\eta^{\prime}$
modes can be enhanced by one order of magnitude.
Recently, the BaBar Collaboration \cite{Babar_ICHEP06}
has measured the semileptonic decays
with the data as follows:
\begin{eqnarray}
BR(B^{+}\to \eta \ell^{+} \nu_{\ell} )&=&(0.84\pm 0.27 \pm
0.21)\times
10^{-4}< 1.4\times 10^{-4} (\rm 90\%\ C.L.)\, ,\nonumber \\
BR(B^{+}\to \eta^{\prime} \ell^{+} \nu_{\ell} )&=&(0.33\pm 0.60 \pm
0.30)\times 10^{-4}< 1.3\times 10^{-4} (\rm 90\%\ C.L.)\, .
\label{Data}
\end{eqnarray}
Although the measurements in Eq. (\ref{Data}) are only
$2.55\sigma$ and $0.95\sigma$
significances, respectively, it is important to examine
if they give some constraints on the form factors due to
the flavor-singlet state in the decays of $\bar B\to
\eta^{(\prime)}\ell \bar\nu_{\ell}$.
It should be also interesting to investigate the implication of the measurements in Eq. (\ref{Data})
 by concentrating on
the flavor-singlet contributions on the flavor changing neutral
current (FCNC) decays of
 $\bar B_{d,s} \to \eta^{(\prime)} \ell^{+} \ell^{-}$.

We start by writing
the effective Hamiltonians for $B^{-}\to
\eta^{(\prime)}\ell \nu_{\ell}$ and $\bar B\to \eta^{(\prime)}\ell^{+} \ell^{-}$ at quark level in the SM as
\begin{eqnarray}
{\cal H}_{I}&=& \frac{G_FV_{ub}}{\sqrt{2}}  \bar u\gamma_{\mu}
(1-\gamma_5) b\, \bar \ell \gamma^{\mu} (1-\gamma_5) \nu_{\ell}\,,
\label{eq:heff_lnu}
\\
 {\cal H}_{II} &=& \frac{G_F\alpha_{em} \lambda^{q^{\prime}}_t}{\sqrt{2}
 \pi}\left[ H_{1\mu} L^{\mu} +H_{2\mu}L^{5\mu}  \right]\,,
 \label{eq:heff_ll}
  \end{eqnarray}
respectively,
  with
  \begin{eqnarray}
  H_{1\mu } &=&C^{\rm eff}_{9}(\mu )\bar{q}^{\prime}\gamma _{\mu }P_{L}b\ -\frac{2m_{b}}{%
 q^{2}}C_{7}(\mu )\bar{q}^{\prime}i\sigma _{\mu \nu }q^{\nu }P_{R}b \,,
\nonumber \\
 H_{2\mu } &=&C_{10}\bar{q}^{\prime}\gamma _{\mu }P_{L}b \,,
 \nonumber\\
 L^{\mu } &=&\bar{\ell}\gamma ^{\mu }\ell\,, \ \ \ L^{5\mu } =\bar{\ell}\gamma ^{\mu }\gamma
 _{5}\ell\,,
  \label{heffc}
  \end{eqnarray}
where $\alpha_{em}$ is the fine structure constant,
$\lambda^{q^{\prime}}_t=V_{tb}V_{tq^{\prime}}^*$, $m_b$ is the
current b-quark mass, $q$ is the momentum transfer, $P_{L(R)}=(1\mp
\gamma_5)/2$  and $C_{i}$ are the Wilson
coefficients (WCs) with their explicit expressions given in
Ref.~\cite{BBL}.
 In particular,  $C^{\rm eff}_{9}$,
 which contains the contribution from the on-shell charm-loop,  is given by
 \cite{BBL}
\begin{eqnarray}
C_{9}^{\rm eff}(\mu)&=&C_{9}( \mu ) +\left( 3C_{1}\left( \mu \right)
+C_{2}\left( \mu \right) \right)  h\left( z,s^{\prime}\right)
\,, \nonumber \\
h(z,s^{\prime})&=&-\frac{8}{9}\ln\frac{m_b}{\mu}-\frac{8}{9}\ln z
+\frac{8}{27} +\frac{4}{9}x  -\frac{2}{9}(2+x)|1-x|^{1/2} \nonumber
\\
&\times& \left\{
  \begin{array}{c}
    \ln \left|\frac{\sqrt{1-x}+1}{\sqrt{1-x}-1} \right|-i\, \pi, \  {\rm for}\ x\equiv 4z^2/s^{\prime}<1 \, , \\
    2\, arctan\frac{1}{\sqrt{x-1}},\   {\rm for}\ x\equiv 4z^2/s^{\prime}>1   \, ,\\
  \end{array}
\right.
\label{C9eff}
\end{eqnarray}
where $h(z,s^{\prime})$ describes the one-loop matrix elements of
operators $O_{1}= \bar{s}_{\alpha }\gamma ^{\mu }P_{L}b_{\beta }\
\bar{c}_{\beta }\gamma _{\mu }P_{L}c_{\alpha }$ and
$O_{2}=\bar{s}\gamma ^{\mu }P_{L}b\ \bar{c}\gamma _{\mu }P_{L}c$
\cite{BBL} with $z=m_c/m_b$ and $s^{\prime}=q^2/m^2_b$.
Here, we have ignored the resonant
contributions \cite{Res,CGPRD66} as they are irrelevant to our analysis.
In Table~\ref{tab:wcs}, we show the values of dominant WCs at
$\mu=4.4$ GeV in the next-to-leading-logarithmic (NLL).
We note that since the value of $|h(z,s^{\prime})|$ is less than 2,
the influence of the charm-loop is much less than $C_{9, 10}$ which
are dominated by the top-quark contributions.
 \begin{table}[hptb]
\caption{
WCs at $\mu=4.4$ GeV in the NLL
order. }\label{tab:wcs}
\begin{ruledtabular}
\begin{tabular}{ccccc}
$C_1$ & $C_2$ & $C_7$ & $C_9$ & $C_{10}$ \\ \hline 
$-0.226$ & $1.096$ & $-0.305$ & $4.344$ & $-4.599$ \\ %
 \end{tabular}
\end{ruledtabular}
\end{table}

To study the exclusive semileptonic decays, the hadronic QCD effects
for the $\bar B\to P$ transitions
are parametrized by
\begin{eqnarray}
\langle P(p_{P}) | \bar q^{\prime} \gamma^{\mu}  b| \bar
B(p_B)\rangle &=& f^{P}_{+}(q^2)\left(P^{\mu}-\frac{P\cdot
q}{q^2}q^{\mu} \right)+f^{P}_{0}(q^2) \frac{P\cdot q}{q^2} q_{\mu}\,
, \nonumber
\\
\langle P(p_{P} )| \bar q^{\prime} i\sigma_{\mu\nu} q^{\nu}b| \bar B
(p_{B})\rangle &=& {f^{P}_{T}(q^2)\over m_{B}+m_{P}}\left[P\cdot q\,
q_{\mu}-q^{2}P_{\mu}\right]\,, \label{eq:bpff}
\end{eqnarray}
with $P_{\mu}=(p_{B}+p_{P})_{\mu}$ and
$q_{\mu}=(p_{B}-p_{P})_{\mu}$. Consequently, the transition
amplitudes associated with the interactions in
Eqs.~(\ref{eq:heff_lnu}) and (\ref{eq:heff_ll}) can be written as
 \begin{eqnarray}
       {\cal M}_{I}&=&\frac{\sqrt{2}G_{F}V_{ub}}{\pi }
        f^{P}_{+}(q^2) \bar{\ell} \not{p}_{P} \ell\,,
       \label{amppln}
       \\
       {\cal M}_{II}&=&\frac{G_{F}\alpha_{em} \lambda^{q^{\prime}} _{t}}{\sqrt{2}\pi }
       \left[ \tm_{97} \bar{\ell} \not{p}_P \ell + \tm_{10} \bar{\ell} \not{p}_P \gamma_5 \ell
        \right]\,,\label{amppll}
 \end{eqnarray}
for $\bar B\rightarrow P \ell \bar\nu_{\ell}$ and
  $\bar B\rightarrow P \ell^{+} \ell^{-}$, respectively,
 with
 \begin{eqnarray}
  \tm_{97}&=& C^{\rm eff}_9 f^{P}_+(q^2) +\frac{2m_b}{m_B+m_{P}}C_7
  f^{P}_T(q^2) \,, \ \ \  \tm_{10}= C_{10} f^{P}_+(q^2)\, .
  \label{eq:m7910}
 \end{eqnarray}
  Since we concentrate
on the productions of the light leptons, we have neglected the terms
explicitly related to $m_{\ell}$. By choosing the coordinates for
various particles:
\begin{eqnarray}
q&=&(\sqrt{q^2},\,0,\, 0,\, 0), \ \ \ p_{B}=(E_{B},\, 0,\, 0,\, |\vec{p}_{P}|), \nonumber \\
p_{P}&=& (E_{P},\, 0,\, 0,\, |\vec{p}_{P}|), \ \ \
p_{\ell}=E_{\ell}(1,\,  \sin\thl,\, 0,\, \cos\thl)\,,
\label{eq:coordinates}
\end{eqnarray}
where $E_{P}=(m^{2}_{B}-q^2-m^2_{P})/(2\sqrt{q^2})$,
$|\vec{p}_{P}|=\sqrt{E^2_{P}-m^2_{P}}$ and
$\theta_{\ell}$ is the polar angle,
the differential decay rates for $ B^{-}\to P \ell
\bar\nu_{\ell}$ and $\bar
B_{d} \to P \ell^+ \ell^-$
as  functions of $q^2$ are given by
\begin{eqnarray}
\frac{d\Gamma_{I}}{dq^2 }&=& \frac{G^{2}_{F} |V_{ub}|^2
m^3_{B}}{3\cdot 2^6 \pi^3}\sqrt{(1-s+\hmp^2)^2-4\hmp^2}
\left(f^{P}_{+}(q^2) \hat P_{P}\right)^2\,,
 \label{eq:diffplnu}
 \\
\frac{d\Gamma _{II} }{dq^2
}&=&\frac{G_{F}^{2}\alpha^{2}_{em}m^{3}_{B}}{ 3\cdot 2^{9} \pi ^{5}}
|\lambda^{q^{\prime}} _{t}|^{2}\sqrt{(1-s+\hmp^2)^2-4\hmp^2} \hat
P^2_{P} \left( |\tilde{m}_{97}|^2+|\tilde{m}_{10}|^2\right) \label{eq:difpll}\, ,
 \end{eqnarray}
respectively, with $\hat P_{P}=2\sqrt{s}
|\vec{p}_{P}|/m_{B}=\sqrt{(1-s-\hmp^2 )^2-4s\hmp^2}$,
$\hat{m}_{P}=m_{P}/m_B$ and $s=q^2/m^2_{B}$.

To discuss the
$P=\eta^{(\prime)}$ modes,
we employ the quark-flavor scheme to describe the states
$\eta$ and $\eta^{\prime}$, expressed by \cite{flavor0,flavor}
\begin{eqnarray}
\left( {\begin{array}{*{20}c}
   \eta   \\
   {\eta '}  \\
\end{array}} \right) = \left( {\begin{array}{*{20}c}
   {\cos \phi } & { - \sin \phi }  \\
   {\sin \phi } & {\cos \phi }  \\
\end{array}} \right)\left( {\begin{array}{*{20}c}
   {\eta _{q} }  \\
   {\eta _{s} }  \\
\end{array}} \right) \label{eq:flavor}
\end{eqnarray}
with $\eta _{q}  = ( {u\bar u + d\bar d})/\sqrt{2}$, $\eta_{s} =
s\bar s $ and $\phi=39.3^{\circ}\pm 1.0^{\circ}$.
Based on this
scheme, it is found that the form factors
in Eq. (\ref{eq:bpff})
at $q^2=0$ with the flavor-singlet contributions
are given by \cite{BN_NPB651}
\begin{eqnarray}
f^{\eta}_{i}(0)&=&\frac{\cos\phi}{\sqrt{2}} \frac{f_{q}}{f_{\pi}}
f^{\pi}_{i}(0) + \frac{1}{\sqrt{3}} \left( \sqrt{2} \cos\phi\frac{
f_{q}}{f_{\pi}} -  \sin\phi  \frac{f_{s}}{f_{\pi}}\right) f^{\rm
sing}_{i}(0)\, ,\nonumber \\
f^{\eta^{\prime}}_{i}(0)&=&\frac{\sin\phi}{\sqrt{2}}
\frac{f_{q}}{f_{\pi}} f^{\pi}_{i}(0) + \frac{1}{\sqrt{3}} \left(
\sqrt{2} \sin\phi\frac{ f_{q}}{f_{\pi}} +  \cos\phi
\frac{f_{s}}{f_{\pi}}\right) f^{\rm sing}_{i}(0)\,, \label{eq:fs}
\end{eqnarray}
where $i=+,T$, $f_{q}=(1.07\pm 0.02)f_{\pi}$, $f_{s}=(1.34\pm 0.06)f_{\pi}$
\cite{flavor} and $f^{\rm sing}_{i}(0)$ denote the
unknown
transition form factors in the
 flavor-singlet mechanism.
 We note that
 the flavor-singlet contributions to $\bar B\to\eta^{(\prime)}$
have also been considered in the soft collinear effective theory \cite{SCET}.
For the $q^2$-dependence form factors $f^{\pi}_{+(T)}(q^2)$,
we quote the results calculated by the light-cone sum rules (LCSR)
\cite{BZ_PRD71}, given by
\begin{eqnarray}
f^{\pi}_{+(T)}(q^2)&=&
\frac{f^{\pi}_{+(T)}(0)}{(1-q^2/m^2_{B^*})(1-\alpha_{+(T)}
q^2/m^2_{B^*})}
\end{eqnarray}
with $f^{\pi}_{+(T)}(0)=0.27$, $\alpha_{+(T)}=0.52(0.84)$ and
$m_{B^*}=5.32$ GeV. Since $f^{\rm sing}_{+,T}(q^2)$ are unknown, as
usual, we parametrize them to be \cite{KOY}
%
\begin{eqnarray}
f^{\rm sing}_{+(T)}(q^2)&=& \frac{f^{\rm
sing}_{+(T)}(0)}{(1-q^2/m^2_{B^*})(1-\beta_{+(T)} q^2/m^2_{B^*})}
\label{eq:ffsing}
\end{eqnarray}
with $\beta_{+(T)}$ being the free parameters. We will demonstrate
that the BRs for the semileptonic decays are not sensitive to the
values of $\beta_{+(T)}$, but those of $f^{\rm
sing}_{+(T)}(0)$.
Moreover,
based on the result of $f^{\pi}_{+}(0)\sim  f^{\pi}_{T}(0)$ in the
large energy effective theory (LEET) \cite{LEET}, we may relate the
singlet form factors of $f^{\rm sing}_{+}(0)$ and $f^{\rm
sing}_{T}(0)$.
Explicitly, we assume that $f^{\rm sing}_{T}(0)\sim  f^{\rm
sing}_{+}(0)$.
 Note that this assumption will not
make a large deviation from the real case since the effects of
$f^{P}_{T}(q^2)$ on the dilepton decays are small due to $C_9>>C_7$
in Eq.~(\ref{eq:m7910}). Hence, the value of $f^{\rm sing}_{+}(0)$
could be constrained by the decays $\bar B\to \eta^{(\prime)} \ell
\bar\nu_{\ell}$.

\begin{figure}[htbp]
\includegraphics*[width=4.5 in]{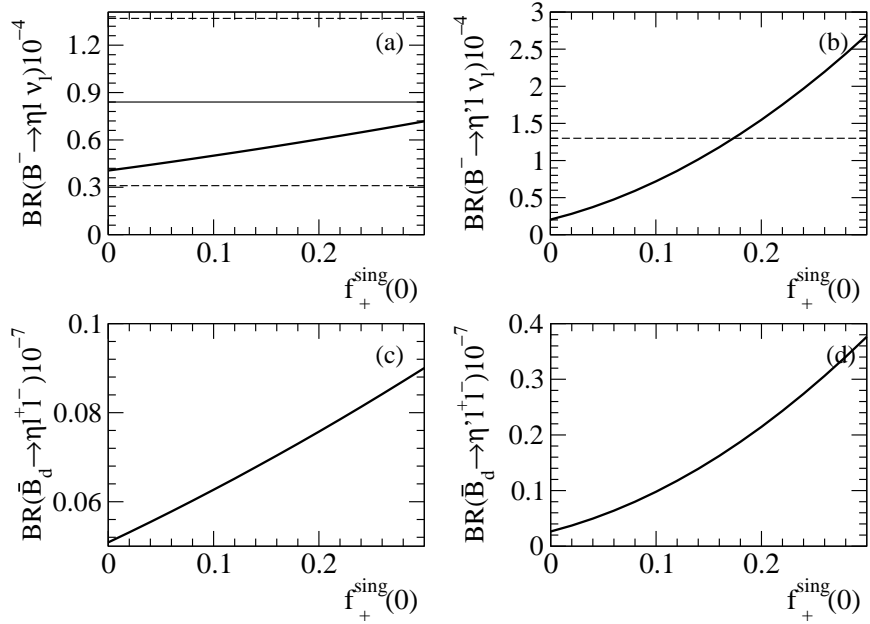}
\caption{BRs of (a)[(b)] $B^{-}\to \eta^{[\prime]} \ell
\bar\nu_{\ell}$ (in units of $10^{-4}$) and (c)[(d)] $\bar B_{d}\to
\eta^{[\prime]} \ell^{+} \ell^{-}$ (in units of $10^{-7}$) as
functions of $f^{\rm sing}_{+}(0)$, where the horizontal solid
 and dashed lines in (a) denote the central and upper and lower values
 of the current
data at $90\%$ C.L., while the dashed line in (b) is the upper limit of
the data.}
 \label{fig:beta}
\end{figure}

Before studying the effects of $f^{\rm sing}_{+}(q^2)$ on the BRs of
semileptonic decays,  we examine the $\beta_{+}$ dependence on the
BRs.
By taking $|V_{ub}|=3.67\times 10^{-3}$
and Eqs.~(\ref{eq:diffplnu}) and (\ref{eq:ffsing}), in
Table~\ref{table:beta}, we present
$BR(B^{-}\to \eta^{(\prime)} \ell\bar\nu_{\ell})$  and $BR(B_{d}\to
\eta^{(\prime)} \ell^{+} \ell^{-})$ with $f^{\rm sing}_{+}(0)=0.2$
and various values of $\beta_{+}$.
 From the table, we see clearly that the
errors of BRs induced by the  errors of  $60\%$ in $\beta_{+}$ are
less than $7\%$ and $14\%$ for the $\eta$ and $\eta^{\prime}$ modes, respectively.
Hence, it is a good approximation to take the $q^{2}$
dependence for $f^{\rm sing}_{+}(q^2)$ to be the same as
$f^{\pi}_{+}(q^2)$. Consequently, the essential effect on the BRs
for semileptonic decays is the value of $f^{\rm sing}_{+}(0)$.
 \begin{table}[hptb]
\caption{BRs of
 $B^{-}\to \eta^{(\prime)} \ell \bar
\nu_{\ell}$ ( in units of $10^{-4}$) and $\bar B_{d} \to
\eta^{(\prime)} \ell^{+} \ell^{-}$ ( in units of $10^{-7}$) with
 $f^{\rm
sing}_{+}(0)=0.2$ and
$\beta_{+}=0.2$, $0.4$, $0.5$, $0.6$ and $0.8$, respectively.}\label{table:beta}
\begin{ruledtabular}
\begin{tabular}{ccccc}
$\beta_{+}$ & $B^{-}\to \eta \ell \bar \nu_{\ell}$& $B^{-}\to
\eta^{\prime} \ell \bar \nu_{\ell}$
& $\bar B_{d}\to \eta \ell^{+} \ell^{-}$ & $\bar B_{d}\to \eta^{\prime} \ell^{+} \ell^{-}$ \\ \hline 
$0.2$ & $0.61$ & $1.37 $ & $0.080$ & $0.190$ \\ \hline
$0.4$ & $ 0.62$ & $ 1.46$  & $0.082$ & $0.204$ \\ \hline
$0.5$ & $ 0.63$ & $ 1.52$  & $0.083$ & $0.211$ \\ \hline
$0.6$ & $0.64$ & $1.58$ & $0.085$ &  $0.220$\\ \hline
$0.8$ & $0.67$ & $1.73$ & $0.088$ &  $0.242$\\
 \end{tabular}
\end{ruledtabular}
\end{table}
With $|V_{td}|=8.1\times 10^{-3}$ \cite{PDG06} and Eqs.
(\ref{eq:diffplnu}) and (\ref{eq:difpll}), the decay BRs of
$B^{-}\to\eta^{(\prime)} \ell \bar\nu_{\ell}$ and $\bar
B_{d}\to\eta^{(\prime)} \ell^{+} \ell^{-}$ as functions
 of $f^{\rm sing}_{+}(0)$  are shown in
Fig.~\ref{fig:beta}.
In Table~\ref{table:rate}, we also explicitly display
the BRs with $f^{\rm sing}_{+}(0)=0$, $0.1$ and $0.2$.
 \begin{table}[hptb]
\caption{BRs of
 $B^{-}\to \eta^{(\prime)} \ell \bar
\nu_{\ell}$ ( in units of $10^{-4}$) and $\bar B_{d} \to
\eta^{(\prime)} \ell^{+} \ell^{-}$ ( in units of $10^{-7}$) with
$\phi=39.3^{\circ}$ and $f^{\rm sing}_{+}(0)=0.0$, $0.1$ and
$0.2$.}\label{table:rate}
\begin{ruledtabular}
\begin{tabular}{ccccc}
$f^{\rm sing}_{+}(0)$ & $B^{-}\to \eta \ell \bar \nu_{\ell}$&
$B^{-}\to \eta^{\prime} \ell \bar \nu_{\ell}$
& $\bar B_{d}\to \eta \ell^{+} \ell^{-}$ & $\bar B_{d}\to \eta^{\prime} \ell^{+} \ell^{-}$ \\ \hline 
$0.0$ & $0.41$ & $0.20 $ & $0.06$ & $0.03$ \\ \hline
$0.1$ & $ 0.52$ & $ 0.71$  & $0.07$ & $0.10$ \\ \hline
$0.2$ & $0.63$ & $1.53$ & $0.08$ &  $0.21$\\
 \end{tabular}
\end{ruledtabular}
\end{table}
 From Table~\ref{table:rate}, we find that
without the flavor-singlet effects, the result for
 $BR(B^{-}\to \eta \ell
\bar\nu_{\ell})$ is a factor of 2 smaller than the
central value of the
BaBar data in Eq. (\ref{Data}).
Clearly, if the data shows a correct tendency,
it indicates that there exist some mechanisms, such as the one with
the flavor-singlet state, to enhance
the decay of $\bar B\to \eta$ as illustrated in Table~\ref{table:rate}.
Moreover,
as shown in Fig.~\ref{fig:beta}b and Table~\ref{table:rate},
the decays of $B^{-}\to \eta^{\prime} \ell \nu_{\ell}$
 are very sensitive to  $f^{\rm sing}_{+}(0)$.
In particular,
the current data has constrained
that
\begin{eqnarray}
f^{\rm sing}_{+}(0)&\leq& 0.2.
\label{Limit}
\end{eqnarray}
It is interesting to
note that for $f^{\rm
sing}_{+}(0)=0.2$, $BR(\bar B_{d}\to \eta^{\prime} \ell^{+} \ell^{-})=
0.21\times 10^{-7}$,
which is as large as
 $BR(B^{-}\to \pi^{-} \ell^{+} \ell^{-})$, while
 that of $\bar B_{d}\to \eta \ell^{+} \ell^{-}$ is
 slightly enhanced.
 In addition,
 it is easy to see that the flavor-singlet contributions
 could result in the BRs of the $\eta^{\prime}$ modes to be over than those of the
 $\eta$ ones.

\begin{figure}[htbp]
\includegraphics*[width=4.in]{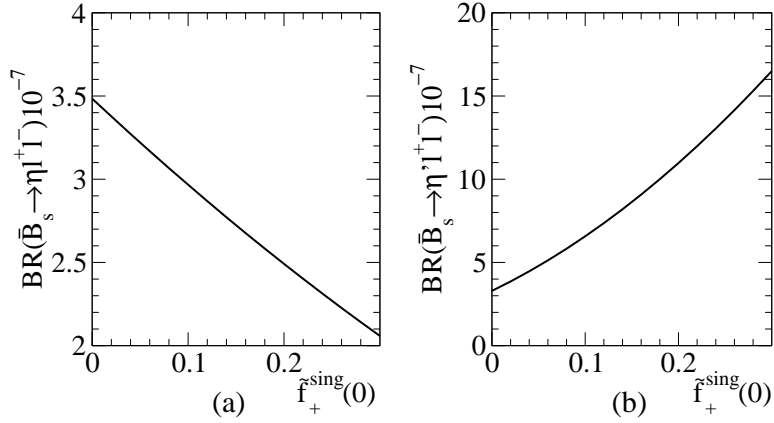}
\caption{(a)[(b)] BRs (in units of $10^{-7}$) of $\bar B_{s}\to
\eta^{[\prime]} \ell^{+} \ell^{-}$  as  functions of
$\tilde{f}^{\rm sing}_{+}(0)$. }
 \label{fig:bseta}
\end{figure}

Our investigation of the
flavor-singlet effects can be extended to the dileptonic decays
of  $\bar B_{s}\to\eta^{(\prime)} \ell^{+} \ell^{-}$ \cite{GL}.
In the following,
we use the notation
with a tilde at the top to represent the form factors associated
with $B_{s}$. Hence,
similar to Eq.~(\ref{eq:fs}), we express the form factors
for $\bar{B}_{s}\to\eta^{(\prime)}$ with
the flavor-singlet effects at $q^{2}=0$ to be
\begin{eqnarray}
\tilde{f}^{\eta}_{+}(0)&=&-\sin\phi
\tilde{f}^{\eta_s[m_{\eta}]}_{+}(0) + \frac{1}{\sqrt{3}} \left(
\sqrt{2} \cos\phi\frac{ f_{q}}{f_{K}} - \sin\phi
\frac{f_{s}}{f_{K}}\right) \tilde{f}^{\rm
sing}_{+}(0)\, ,\nonumber \\
\tilde{f}^{\eta^{\prime}}_{+}(0)&=&\cos\phi
\tilde{f}^{\eta_s[m_{\eta^{\prime}}]}_{+}(0) + \frac{1}{\sqrt{3}}
\left( \sqrt{2} \sin\phi\frac{ f_{q}}{f_{K}} + \cos\phi
\frac{f_{s}}{f_{K}}\right) \tilde{f}^{\rm sing}_{+}(0)\,.
\end{eqnarray}
 For the $q^2$-dependence
form factors
of $\tilde{f}^{\eta_{s}[m_{\eta^{(\prime)}}]}_{+,T}$,
we adopt the results calculated by the constituent
quark model (CQM) \cite{CQM},
given  by
\begin{eqnarray}
\tilde{f}^{\eta_{s}[m_{\eta^{(\prime)}}]}_{+,T}(q^{2})=\frac{\tilde{f}^{\eta_{s}[m_{\eta^{(\prime)}}]}_{+,T}(0)}{1-a^{\eta^{(\prime)}}_{+,T}
q^2/m^{2}_{B_s}+b^{\eta^{(\prime)}}_{+,T} (q^2/m^2_{B_s})^2}
\end{eqnarray}
with
$\tilde{f}^{\eta_{s}[m_{\eta}]}_{+}(0)=\tilde{f}^{\eta_{s}[m_{\eta^{\prime}}]}_{+}(0)=\tilde{f}^{\eta_s[m_{\eta}]}_{T}(0)=0.36$,
$\tilde{f}^{\eta_s[m_{\eta^{\prime}}]}_{T}(0)=0.39$,
$a^{\eta}_{+}=a^{\eta^{\prime}}_{+}=0.60$,
$b^{\eta}_{+}=b^{\eta^{\prime}}_{+}=0.20$,
$a^{\eta}_{T}=a^{\eta^{\prime}}_{T}=0.58$ and
$b^{\eta}_{T}=b^{\eta^{\prime}}_{T}=0.18$.
By using $m_{B_s}=5.37$
GeV and  $V_{ts}=-0.04$ instead of $m_{B}$ and $V_{td}$ in
Eq.~(\ref{eq:difpll}), we present the BRs of $\bar B_{s}\to
\eta^{(\prime)} \ell^{+} \ell^{-}$ in Table~\ref{table:ratebs}.
We also display the BRs as  functions of $\tilde{f}^{\rm
sing}_{+}(0)$ in Fig.~\ref{fig:bseta}.
\begin{table}[hptb]
\caption{BRs of $\bar B_{s}\to \eta^{(\prime)} \ell^{+} \ell^{-}$ (
in units of $10^{-7}$) with $\phi=39.3^{\circ}$ and $\tilde{f}^{\rm
sing}_{+}(0)=0.0$, $0.1$ and $0.2$.}\label{table:ratebs}
\begin{ruledtabular}
\begin{tabular}{cccc}
$\tilde{f}^{\rm sing}_{+}(0)$ & $0.0$& $0.1$ & $0.2$ \\ \hline 
$\bar B_{s}\to \eta \ell^{+} \ell^{-}$ &  $3.71$ & $3.27$ &  $2.84$  \\
\hline
$\bar B_{s} \to \eta^{\prime} \ell^{+} \ell^{-}$ & $3.35$ & $5.97$ &  $9.35$  \\
 \end{tabular}
\end{ruledtabular}
\end{table}
%
As seen from Table~\ref{table:ratebs} and Fig.~\ref{fig:bseta}, due to the
flavor-singlet effects,
the BRs of $\bar B_{s} \to \eta^{\prime}\ell^{+}
\ell^{-}$ are enhanced and
could be as large as $O(10^{-6})$ with around a factor of $3$
enhancement,
whereas those of $\bar B_{s} \to \eta\ell^{+}\ell^{-}$ decrease
as increasing $\tilde{f}^{\rm sing}_{+}(0)$,
which can be tested in future hadron colliders.

In summary, we have studied the effects of the
flavor-singlet state on the
$\eta^{(\prime)}$ productions in the semileptonic B decays. In terms of
the constraints from the
current data of $B^{-}\to \eta^{(\prime)} \ell
\nu_{\ell}$, we have found that the BRs of $\bar B_{d,s}\to
\eta^{\prime}\ell^{+} \ell^{-}$ could be enhanced to be $O(10^{-8})$
and $O(10^{-6})$, respectively.
Finally, we remark that
the flavor-singlet effects could result in
the BRs of
the $\bar B_{d,s}\to
\eta^{\prime}$ modes to be larger than those of $\bar B\to \eta$
and the statement
is reversed if the effects are neglected.

Note added: After we presented the paper, Charng, Kurimoto and Li
\cite{Li} calculated the flavor singlet contribution to the
$B\to \eta^{(')}$ transition form factors from the gluonic content of
$\eta^{(')}$ in the large recoil region by using the perturbative QCD
(PQCD) approach.
Here, we make some comparisons as follows:\\
1.
While Ref. \cite{Li} gives a theoretical calculation on the flavor
singlet contribution to the form factors in the PQCD, we consider
the direct constraint from the experimental data.
The conclusion that the singlet contribution is negligible (large)
in the $B\to \eta^{(')}$ form factors in Ref. \cite{Li} is the same as
ours. However, the overall ratios in Ref. \cite{Li} for the singlet contributions are about a factor 4 smaller than ours.
On the other hand, as stressed in Ref. \cite{Li}, they have used a small
Gegenbauer coefficient, which corresponds smaller gluonic contributions.
For a larger allowed value of the the Gegenbauer coefficient in
Ref. \cite{BN_NPB651}, the overall ratios
will be a few factors larger. In other words, the real gluonic contributions  rely on
future experimental measurements.
\\
2. Although our assumption of $f^{\rm sing}_{T}(0)\sim f^{\rm
sing}_{+}(0)$ seems to be
somewhat different from the PQCD calculation as pointed out in Ref.
\cite{Li} due to an additional term,
the numerical values of
$f^{\rm sing}_{+}(0)=0.042$ and $f^{\rm sing}_{T}(0)=0.035$ by
the PQCD \cite{Li}
do not change our assumption very much. After all, as stated in point 1 that there
exist large uncertainties for the wave functions in the PQCD.
In addition, the
difference actually is not important for our results as
explained in the text.
\\
\\
\\
{\bf Acknowledgments}\\

This work is supported in part by the National Science Council of
R.O.C. under Grant \#s:NSC-95-2112-M-006-013-MY2,
NSC-94-2112-M-007-(004,005) and NSC-95-2112-M-007-059-MY3.


\begin{thebibliography}{99}

\bibitem{belle_0603} BELLE Collaboration, K. Abe {\it et al.},
arXiv:hep-ex/0603001.

\bibitem{babar_0608} BABAR Collaboration, B. Aubert {\it et al.},
Phys. Rev. Lett. {\bf 94}, 191802 (2005); arXiv:hep-ex/0608005.

\bibitem{Kroll}
P.~Kroll and K.~Passek-Kumericki,
  Phys.\ Rev.\ D {\bf 67}, 054017 (2003)
  [arXiv:hep-ph/0210045].

\bibitem{BN_NPB651} M. Beneke and M. Neubert, Nucl. Phys. B{\bf
651}, 225 (2003) [arXiv:hep-ph/0210085].

\bibitem{KOY}C.S. Kim, S. Oh and C. Yu, Phys. Lett. B{\bf 590}, 223
(2004) [arXiv:hep-ph/0305032].

\bibitem{Babar_ICHEP06} Babar Collaboration, B. Aubert {\it et al.},
arXiv:hep-ex/0607066.

\bibitem{BBL}  G. Buchalla, A. J. Buras and M. E. Lautenbacher, Rev.
Mod. Phys {\bf 68}, 1230 (1996) [arXiv:hep-ph/9512380].

\bibitem{Res}
 C.~S.~Lim, T.~Morozumi and A.~I.~Sanda,
  Phys.\ Lett.\ B {\bf 218}, 343 (1989);
  N.~G.~Deshpande, J.~Trampetic and K.~Panose,
  Phys.\ Rev.\ D {\bf 39}, 1461 (1989);
P.~J.~O'Donnell and H.~K.~K.~Tung,
  Phys.\ Rev.\ D {\bf 43}, 2067 (1991).

\bibitem{CGPRD66} C.H. Chen and C.Q. Geng, Phys. Rev. D{\bf 66}, 034006
(2002) [arXiv:hep-ph/0207038].

\bibitem{flavor0}
 J.~Schechter, A.~Subbaraman and H.~Weigel,
  Phys.\ Rev.\ D {\bf 48}, 339 (1993).

\bibitem{flavor}
T. Feldmann, P. Kroll and B. Stech, Phys. Rev. D{\bf 58},
114006(1998) [arXive:hep-ph/9802409].

\bibitem{SCET}
 A.~R.~Williamson and J.~Zupan,
  Phys.\ Rev.\ D {\bf 74}, 014003 (2006)
  [Erratum-ibid.\ D {\bf 74}, 03901 (2006)]
  [arXiv:hep-ph/0601214].

\bibitem{BZ_PRD71}P. Ball and R. Zwicky, Phys. Rev. D{\bf 71}, 014015 (2005) [arXiv:hep-ph/0406232];
Phys. Lett. B{\bf 625}, 225 (2005) [arXiv:hep-ph/0507076].

\bibitem{LEET} M.J. Dugan and B. Grinstein, Phys. Lett. B{\bf 255}, 583
(1991); J. Charles {\it et. al.}, Phys. Rev. D{\bf 60}, 014001
(1999).

\bibitem{PDG06}Particle Data Group, W.M. Yao {\it et al.}, J. Phys. G {\bf 33}, 1 (2006).


\bibitem{GL} C.Q. Geng and C.C. Liu,
J.\ Phys.\ G {\bf 29}, 1103 (2003)
  [arXiv:hep-ph/0303246].


\bibitem{CQM} D. Melikhov and B. Stech, Phys. Rev. D{\bf 62},
014006 (2000) [arXiv:hep-ph/0001113].

\bibitem{Li}
 Y.~Y.~Charng, T.~Kurimoto and H.~n.~Li,
  Phys.\ Rev.\ D {\bf 74}, 074024 (2006)
  [arXiv:hep-ph/0609165].

\end{thebibliography}
\end{document}